\documentclass[psfig,epsfig]{aastex}
\usepackage{emulateapj5}
\usepackage{apjfonts}

\received{}
\accepted{}
\newcommand{\msolar}{\mbox{\,$M_{\odot}$}}

\newcommand{\temp}{{\rm T}}
\newcommand{\ovi}{{\rm O{\textsc{vi}}\,}}

\slugcomment{Submitted to the Astrophysical Journal (Letters), 20 May 2006}
\lefthead{Avillez \& Breitschwerdt}
\righthead{Supernova-driven Interstellar Turbulence}
\begin{document}
\title{The Generation and Dissipation of Interstellar
Turbulence - Results from Large Scale High Resolution Simulations}
\author{Miguel A. de Avillez$^{1,2}$ and Dieter Breitschwerdt$^{2}$}
\affil{$^{1}$Department of Mathematics, University of \'Evora, R. Rom\~ao
Ramalho 59, 7000 \'Evora, Portugal\\
$^{2}$ Institut f\"ur Astronomie, University of Vienna,
T\"urkenschanzstra{\ss}e 17, A-1180 Vienna, Austria\\
mavillez,~breitschwerdt@astro.univie.ac.at}

\begin{abstract}
We study, by means of adaptive mesh refinement hydro- and
magnetohydrodynamical simulations that cover a wide range of scales
(from kpc to sub-parsec), the dimension of the most dissipative
structures and the injection scale of the turbulent interstellar gas,
which we find to be about 75 pc, in agreement with observations. This
is however smaller than the average size of superbubbles, but
consistent with significant density and pressure changes in the ISM,
which leads to the break-up of bubbles locally and hence to injection
of turbulence. The scalings of the structure functions are consistent
with log-Poisson statistics of supersonic turbulence where energy is
dissipated mainly through shocks. Our simulations are different from
previous ones by other authors as (i) we do not assume an isothermal
gas, but have temperature variations of several orders of magnitude
and (ii) we have no artificial forcing of the fluid with some ad hoc
Fourier spectrum, but drive turbulence by stellar explosions at the
Galactic rate, self-regulated by density and temperature thresholds
imposed on the ISM gas.
\end{abstract}


\section{Introduction}

Long ago it was recognized that the interstellar gas is a highly
turbulent and compressible medium (von Weizs\"acker 1951), although
this fact is still being ignored in many theoretical models. Indeed,
high resolution observations of the ISM show structures on \emph{all
scales} down to the smallest resolvable ones, implying a dynamical
coupling over a wide range of scales, which is a main characteristic
of a turbulent flow with Reynolds numbers the order of $Re = 10^5 -
10^7$ (cf.\ Elmegreen \& Scalo 2004). It was not until the last decade
that the effects of turbulence in the ISM in general (e.g.,
V\'azquez-Semadeni et al. 1995, 2000; Passot et al. 1995; Korpi et
al. 1999; Wada \& Norman 2001; Avillez \& Mac Low 2002; Avillez \&
Breitschwerdt 2004, 2005 - herafter referred to as AB04, AB05; Joung
\& Mac Low 2006) and in molecular clouds, in particular (e.g., Mac Low
et al. 1998; Stone et al. 1998; Porter et al. 1998, 1999; Padoan \&
Nordlund 1999; Boldyrev et al. 2002; Padoan et al. 2004), have been
studied with high detail by means of numerical simulations. The
molecular cloud studies focused on decaying and continuously
driven turbulence by means of large-scale incompressible forcing in
Fourier space, stirring the gas in the entire computational domain
simultaneously. From these simulations it became clear that (i)
isothermal supersonic turbulent motions inside molecular clouds decay
within a sound crossing time becoming subsonic whether a magnetic
field is present or not (Mac Low et al. 1998; Stone et al. 1998;
Padoan \& Nordlund 1999) and (ii) the fractal dimension of the most
dissipative structures in molecular clouds evolves from close to 1
(filaments) to approximately 2 (shocks) with increasing Mach number
(Padoan et al. 2004).

Although there has been a lot of observational, theoretical and
computational research concerning interstellar turbulence, little is
known about: (i) the scales at which energy is injected, and
(ii) the dimension of the most dissipative structures, if a wide range
of scales (from kpc to sub-parsec) is taken into account and the
turbulence forcing are point explosions acting in small regions of the
domain as opposed to incompressible forcing over the entire domain.

Here we address these issues by means of 3D adaptive mesh refinement
(AMR) hydro- (HD) and magnetohydrodynamical (MHD) simulations of the
SN-driven ISM.  The outline of the present paper is as follows:
Section~2 deals with the physical model, setup of the
simulations and numerical code used; In Section~3 a characterisation
of the source and injection scale of the interstellar turbulence is
made, while in Section 4 the Hausdorff dimension of the most
dissipative structures is analysed; Section~5 closes the paper with a
discussion and conclusions.

\section{Model and Simulations}

We use previously published (AB04, AB05) 3D AMR HD and MHD simulations
(denoted by HD04 and MHD05) and a new 0.625 pc HD run (denoted by
HD06) with self-gravity of the SN-driven ISM in a section of the
Galaxy centered at the Solar circle having an area of 1 kpc$^{2}$ and
extending up to 10 kpc on either side of the Galactic midplane,
capturing both the large scale disk-halo-disk circulation as well as
the smallest structures in the Galactic disk. All the runs use a 10 pc
resolution coarse grid and generate subgrids up to a local resolution
of 1.25 pc (3 levels of refinement; HD04 and MHD05 runs) and 0.625 pc
(4 levels of refinement; HD06 run) in the disk region
$\left|z\right|\leq 500$ pc (corresponding to effective grids of
$800^{3}$ and $1600^{3}$ cells, respectively) and 2.5 pc (2 levels of
refinement) on the remaining computational domain. The boundary
conditions are periodic along the vertical faces and outflow on the
top and bottom faces of the grid.

The physical model, which does not include heat conduction (because we
find that turbulent diffusion is a significantly more efficient
transport process), includes a gravitational field provided by the
stars in the disk (Kuijken \& Gilmore 1989), local self-gravity
(excluding the contribution from the newly formed stars), radiative
cooling assuming collisional ionization equilibrium (following
Dalgarno \& McCray (1972) and Sutherland \& Dopita (1993) cooling
functions for gas with $10 \leq \temp < 10^{4}$ and $10^{4} \leq \temp
\leq 10^{8.5}$ K, respectively) and solar abundances (Anders \& Grevesse 1989),
uniform heating due to a UV radiation field normalized to the Galactic
value and varying with $z$ and photoelectric heating of grains and
polycyclic aromatic hydrocarbons (Wolfire et al. 1995). The disk 
gas is driven by supernovae types Ia and II (including types Ib, Ic and II) with the observed Galactic SN rates
(Cappellaro et al. 1999). SNe Ia are set up randomly with an
exponential distribution having a scale height of 325 pc (Freeman
1987). The number and masses of SNe II progenitors are determined
according to the local mass distribution in the highest density ($n>
10$ cm$^{-3}$) cold ($\temp\le 100$ K) regions of converging flows
($\nabla\cdot \vec{v}<0$; here $\vec{v}$ is the gas velocity) on
the level 1 grid (note that the coarse grid is identified as
level 0) after synchronization of the time steps of the finer and
coarser grids. In the flagged regions, with a star forming efficiency
of 10\%, the available mass that can be turned into stars is $>840
\msolar$. The masses and number of the new stars are determined by an
initial mass function (IMF) with lower and upper mass cutoffs of 8 and
60 $\msolar$, respectively, in order to guarantee that all the
available mass goes into high mass ($> 8 \msolar$) stars. For the
present work low mass stars which should be formed coevally, are not modelled, but they should have a minor influence (except for the stellar gravitational field, which has already been accounted for) on the ISM dynamics.

The newly formed stars, to which the observed mean random velocity of
5 km/s is added at time of formation, are followed kinematically until
the end of their main sequence life time determined from Stothers'
(1972) formula. The rate at which these stars explode is normalized to
the Galactic rate, that is, if the SN rate in the disk is
larger/smaller by 10\% of the observed value, SN occurrences are
reduced/forced artificially. In general 40-50\% of the type II SN
progenitors explode in the field, while the remaining explode
correlated in time and space generating large superbubble structures,
some of which have sizes as large as 500 pc (see AB04, AB05). 

The simulations follow a single fluid with an initial vertical density
distribution that includes the contributions due to the molecular,
neutral, ionized and hot phases observed in the ISM (Ferri\`ere
2001). The mean magnetic field component in the MHD05 run is
initialized assuming equipartition, while the random component, $\delta \vec{B}$, is zero. During the first 20 Myr $\delta \vec{B}$ builds up and the
total field becomes 4.5 $\mu$G (AB05).

We use the 3D \'Evora-Vienna Astrophysics Fluid-Parallel AMR
(EVAF-PAMR) Code originally developed by Avillez (2000). It is a
Fortran 95 code that solves HD and MHD problems with AMR (Pember et
al. 1996; Balsara 2001) in a parallel fashion and uses approximate
Riemann solvers for the hydro and magnetic components (Collela \&
Woodward 1984; Dai \& Woodward 1994, 1998).

\section{The injection scale of interstellar turbulence}

The simulations were run for 400 Myr, a time sufficiently long to wipe
out any signatures of the initial conditions, which are still present
at 80 Myr of evolution (see AB04). The initially unstable vertical gas
distribution reaches a global dynamical equilibrium at $\sim200$ Myr
after the set-up of the Galactic fountain, with hot gas ascending into
the halo where it cools and finally returns to the disk. During the
dynamical equilibrium evolution the Galactic midplane is filled with a
thin cold layer overlayed by a thick frothy disk composed mainly of warm neutral and ionized gas having scale heights of 180 and 900 pc, resembling closely the Lockman (Lockman et al. 1986) and Reynolds (1987) layers. In the disk the highest density (and lowest temperature, if the gas
had enough time to cool) gas tends to be confined to sheet-like
structures (2D), which are formed by SN driven shock compressed
layers. In the HD runs the $\temp < 10^{3}$ K gas has no preferred
morphology, while in the MHD05 run it tends to be aligned with the
magnetic field. The highest densities in the present simulations can
be up to $10^{4}$ cm$^{-3}$ as a result of the self-gravity occurring
in the cold clouds, which are mainly formed in regions where several
streams of convergent flows meet. The orientation of these streams is
random. We note in passing, that if clouds are hit by shocks from
random directions, turbulence in the interior is generated. Tapping
the turbulent ISM energy reservoir, which is huge, represents a neat
way of sustaining supersonic turbulence inside molecular clouds
against efficient dissipation.

\begin{figure}[thbp]
\centering
\includegraphics[width=0.4\hsize,angle=0]{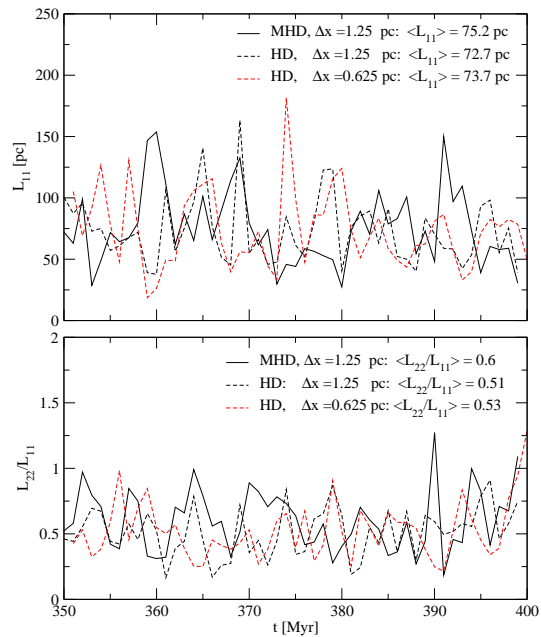}
\caption{History of the characteristic size (given by $L_{11}$) of
the larger eddies (top panel) and of the ratio $L_{22}/L_{11}$ (bottom
panel) for the MHD (solid line) and HD (dashed lines) runs for 1.25 pc
(black) and 0.625 pc (red) resolutions.}
\label{l11}
\end{figure}

The scale at which energy is transferred to the interstellar gas can
be determined by using the so-called two-point correlation function
$R_{ij}(\vec{l},t)=\langle
u_{i}(\vec{x}+\vec{l},t)u_{j}(\vec{x},t)\rangle$, where $u_{i}, u_{j}$
are the components of the fluctuating velocity field $\vec{u}$.  The
diagonal components of $R_{ij}$ are even functions of $\vec{l}$ and
can be written in terms of the dimensionless scalar functions $f(l,t)$
and $g(l,t)$, l=$\|\vec{l}\|$, which satisfy $f(0,t)=g(0,t)=1$ and
$f,~g\leq 1$, as $R_{11}/u^{2}=f(l,t)$ and $R_{ij}/u^{2}=g(l,t)$ if
$i=j\neq 1$ and zero if $i\neq j$ with $u=(\frac{1}{3}\langle
\vec{u}\rangle)^{1/2}$. The characteristic injection scale in the flow is 
given by $L_{11}=\int_{0}^{+\infty}f(l,t) dl$, which in the present
simulations is calculated in a region with a linear size of 500 pc at
a distance of 250 pc from the edges of the computational domain to
avoid the periodicity effects of the boundary conditions. The top
panel of Figure~\ref{l11} shows the history of $L_{11}$ in the last 50
Myr of evolution of the simulated ISM for 1.25 (in black) and 0.625 pc
(in red) resolutions. The time average of $L_{11}$ varies between 73
and 75 pc for the HD and MHD cases, respectively, and seems to be
independent of resolution. In all the cases considered here, there is
a large scatter of $L_{11}$ around its mean as a result of the
expansion of bubbles and superbubbles in an inhomogeneous medium
characterized by a large number of SNe exploding in the field and the
merging of bubbles and superbubbles. This scale corresponds to the one
at which the power spectrum of the solenoidal component of the
velocity field has a spectral break, which we interpret as a
redistribution of energy to both larger and smaller scales (Avillez \&
Breitschwerdt 2007, in preparation).

The ratio between the transverse, $L_{22}=\int_{0}^{+\infty}
g(l,t)dl$, and longitudinal correlation lengths
($L_{22}/L_{11}$, which in isotropic turbulence is 0.5) varies between
0.2 and 1.3 (bottom panel of Figure~\ref{l11}), being its time
average, over a period of 50 Myr, $\langle
L_{22}/L_{11}\rangle_{t}\sim 0.5$ (HD04 and HD06 runs) and 0.6 (MHD05
run). The former value indicates that in a statistical sense the
interstellar unmagnetized turbulence is roughly isotropic, while the
20\% deviation from 0.5 seen in the MHD05 run suggests that $\langle
L_{22}/L_{11}\rangle_{t}$ increases with the magnetic field
strength. This is also supported by similar increases observed in
other MHD runs we carried out (0.25, 0.5, 1, 1.5 and 2 $\mu$G), which
will be described in a forthcoming paper.

\section{Intermittency and Hausdorff Dimensions}

From the Kolmogorov (1941, herafter K41) model for homogeneous and isotropic
turbulence one can derive that, in the inertial range, the moments of order p,
$\langle \delta v_{l}^{p}\rangle$, are dictated solely by $\langle
\delta v_{l}^{3}\rangle$ via the set of scalings $\zeta(p)/\zeta(3)$,
\begin{equation}
\label{scaling1}
\label{eq2}
\displaystyle \langle \delta v_{l}^{p}\rangle\propto \langle \delta
v_{l}^{3}\rangle^{\zeta(p)/\zeta(3)},
\end{equation}
with $p>1$ being an integer. In this expression $\delta v_{l} =v(x+l)-v(x)$ with
$v(x+l)$ and $v(x)$ being the velocities along the $x-$axis at two points
separated by a distance $\eta \ll l\ll L_{11}$, with $L_{11}$ being the
outer scale and $\eta=\nu^{3/4}\epsilon^{-1/4}$ the Kolmogorov
microscale; $\langle\;\rangle$ stands for the ensemble average over the
probability density function of $\delta v_{l}$. Experiments have shown that
(see discussions in Benzi et al. 1993 and Frisch 1995): (i) equation
(\ref{scaling1}), with the same scaling exponents, is valid in a wide range of
length scales for large as well as small $Re$ even if no inertial range is
established and (ii) in the inertial range of high $Re$ flows the scaling
exponents become significantly smaller than the linear law $p/3$ (the K41
scaling) with increasing $p$. This is interpreted as the statistics of small
scales (the ones dictating the behaviour of higher-order moments) becoming
increasingly non-gaussian.

\begin{figure}[thbp]
\centering
\includegraphics[width=0.4\hsize,angle=0]{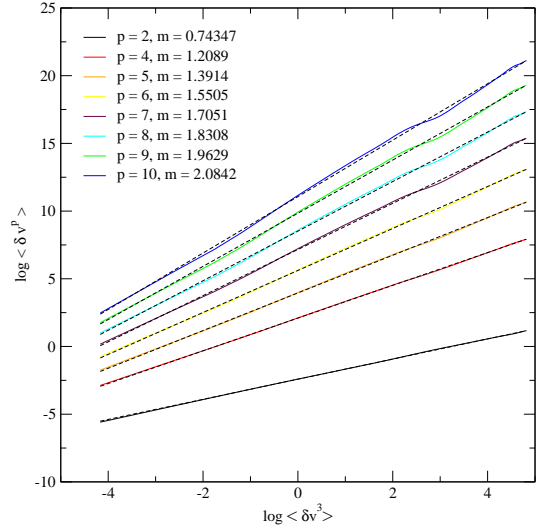}
\caption{Log-log plot of the velocity correlation moments $\langle \delta v_{l}^{p}\rangle$ as function of
$\langle\delta v_{l}^{3}\rangle$ and best fits (dashed lines) for
$p=2,4,...,10$. The data refer to the 0.625 pc resolution run.}
\label{slopes}
\end{figure}

In the unmagnetized and magnetized simulated interstellar turbulence
driven by point explosions the moments $\langle \delta v_{l}^{p}
\rangle$ are related to $\langle \delta v_{l}^{3}\rangle$ (see
Figure~\ref{slopes}, which shows the moments and the slopes
$\zeta(p)/\zeta(3)$ of their best fits for the HD06 run), supporting
the validity of equation (\ref{scaling1}) in such cases of
compressible forcing.

\begin{figure}[thbp]
\centering
\includegraphics[width=0.4\hsize,angle=-90]{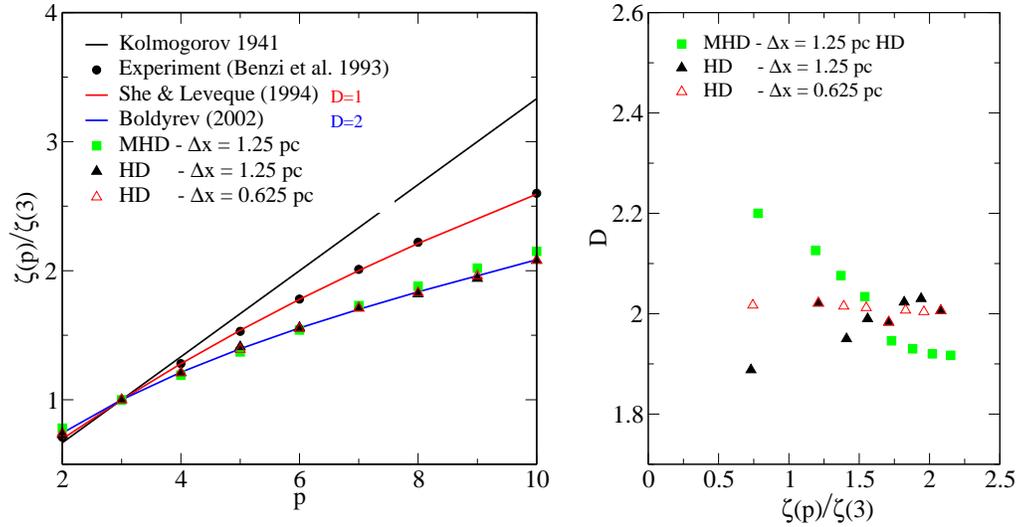}

\caption{\emph{Left:} Exponent $\zeta(p)/\zeta(3)$ for the structure
function versus order $p$. The full line corresponds to
$\zeta(p)=p/3$ (K41 theory), bullets correspond to data of Benzi et
al. (1993), red and blue lines refer to the She-Lev\'eque (1994) and
Burgers-Kolmogorov (Boldyrev 2002) models, respectively. Triangles
represent data from HD simulations with 1.25 pc (black) and 0.625 pc
(red) resolutions and the green squares refer to the MHD run with
1.25 pc resolution. Fractal dimension $D$ of the most
dissipative structures (derived from eq.~\ref{eqscalings}) as a
function of the exponent $\zeta(p)/\zeta(3)$, with $p=2,4,...,10$,
shown in the left panel for the three simulations.
\label{scalings} }
\end{figure}

The left panel of Figure~\ref{scalings} compares theoretical,
experimental and simulated (black and red triangles refer to the HD04
and HD06 runs, respectively, while squares refer to the MHD05 run)
results on the variation of $\zeta(p)/\zeta(3)$ with $p$. The
experimental data is taken from Benzi et al.  (1993) and the
theoretical predictions correspond to the K41 and She-Lev\'eque (1994;
SL94) models of incompressible turbulence and the Burgers-Kolmogorov
model (Boldyrev 2002; BK02) for supersonic turbulence. The K41 model
considers a log-normal statistics for the transfer of energy from
large to small scales and has no corrections for intermittency. The SL94 model uses a log-Poisson statistics to describe intermittency and assumes that energy dissipation occurs through filamentary structures (Hausdorff dimension $D=1$) resulting from vortex stretching. The BK02 and SL94 models differ in the
geometry of the most intermittent structures, which for supersonic
turbulence are assumed to be shocks (Hausdorff dimension $D=2$). We
note that the geometry of those scales does not have to represent
\emph{dissipative structures} (as in incompressible turbulence) nor do
they have to be \emph{planar}. 2D in the Hausdorff sense could
represent various structures, sheets being the simplest.  This is
further supported by the fact that Burger's turbulence produces
discontinuities, but not shocks, in the sense that material cannot
flow through them -- they just sweep up material.

The scalings $\zeta(p)/\zeta(3)$ can be written as (Dubrulle 1994; see
discussion in Politano \& Pouquet 1995)
\begin{equation}
\label{eqscalings}
\frac{\zeta(p)}{\zeta(3)}=\frac{p}{g}(1-x)+(3-D)\left(1-\beta^{p/g}\right),
\end{equation}
where $g$ is linked to the model of nonlinear energy transfer via the basic
scaling $v_{l}\sim l^{1/g}$, $x$ is related to the energy transfer time at the
smallest scale $t_{l}\sim l^{x}$, with $x>0$ and $\beta=1-x/(3-D)$. In these
expressions $D$ is the fractal dimension of the most intermittent structures.
In the present simulations turbulence is mostly solenoidal in the inertial
range and, therefore, $g=3$ and $x=2/3$. Thus, the Hausdorff dimension $D$ of
the most intermittent structures derived from (\ref{eqscalings}) and using the
$\zeta(p)/\zeta(3)$ values shown in the left panel of Figure~\ref{scalings}
varies between 1.9 and 2.2 ($D=1.98-2.02$, $1.9-2.03$ and $1.91-2.2$ in the
HD06, HD04 and MHD05 runs, respectively; right panel of Figure~\ref{scalings}).
These values suggest that the energy injected by supernovae into interstellar
turbulence is dissipated preferentially through 2D structures that can be
identified as shock surfaces in the HD cases and also as current sheets in the
MHD case (see M\"uller \& Biskamp 2000). The discrepancy observed in the MHD
run, although within the errors of numerical noise, indicates a tendency toward
filamentary dissipative structures due to the anisotropy induced by the
magnetic field. We postpone a more detailed discussion on this issue 
to a forthcoming paper. 

\section{Discussion and Conclusions}

Contrary to previous work of other authors, the present runs
incorporate (i) the full extent disk-halo-disk circulation, (ii) do
not assume isothermality (temperature varies within several orders of
magnitude), (iii) include magnetic fields and self-gravity, and, (iv)
cover a wide range (from kpc to sub-parsec) of scales, providing
resolution independent information on the injection scale, extended
self-similarity and dissipative structures. The mean integral scale at
which energy is injected into the interstellar medium is of the order
of 75 pc. This scale is significantly lower than the often quoted
average size of evolved SNRs or superbubbles, because energy is
injected already at the timescale of break-up of the remnants due to
density and pressure gradients in the inhomogeneous ambient medium,
which occurs well before they stall. The similarity between the HD and
MHD results indicates that magnetic pressure and tension forces cannot
prevent break-out of the hot gas from the bubbles formed by SN
explosions, as long as $L\lesssim \beta_{P} \xi$, where $L$ and $\xi$
are the scale lengths of thermal and magnetic pressures (including
tension forces) and $\beta_{P}=4\pi P/B^{2}$ is the plasma
beta. However, the magnetic field introduces a degree of anisotropy
which is reflected in the 20\% deviation from 0.5 seen in the time
averaged ratio $L_{22}/L_{11}$ of the magnetized ISM. Kaplan (1958)
already suggested an injection scale of 80 pc based on the analysis of
second order structure functions of interstellar matter. The
dissipation of energy in our interstellar turbulence simulations
proceeds through shocks and numerical viscosity, as we cannot resolve
the physical viscosity scales. The variation of $\zeta(p)/\zeta(3)$
with $p$ is most consistent with a log-Poisson model for the scales at
which intermittency becomes important.

Recently Joung \& Mac Low (2006) found similar scalings in a SN-driven
ISM HD simulation, although they do not consider self-gravity, find a
hot (T$>10^{5.5}$ K) gas occupation fraction $f_{h}>40\%$, which is at
odds with observations and other models, and their resolution (1.95
pc) in the disk is a factor of 3 lower than that in the HD06 run
(0.625 pc). Their high value of $f_{h}$ implies that the amount of
\ovi in the ISM is a least a factor $2$ larger than that observed with
Copernicus and FUSE. According to our simulations, reliable
comparisons of global properties with observations (e.g., occupation
fractions) can only be made after the system reaches a dynamical
equilibrium ($t>150$ Myr), implying a relaxation time much larger than
a few million years (see discussion in AB04).

\begin{acknowledgments}
We thank the referee and the editor, John Scalo, for detailed reports
that allowed us to significantly improve the paper. This research is
supported by the Portuguese Science Fountation (FCT) through project
POCTI/FIS/58352/2004.
\end{acknowledgments}

\end{document}